\documentclass[prb,aps,floatfix,twocolumn,showpacs,superscriptaddress,amsmath,amssymb]{revtex4}
\usepackage{graphicx}
\usepackage{dcolumn}
\usepackage{color}
\usepackage{pst-all}
\usepackage{bm}
\begin{document}

\title{Excitations with fractional spin less than $\frac12$ in frustrated magnetoelastic chains}
\author{C.J.\ Gazza}
\address{Facultad de Ciencias Exactas Ingenieria y Agrimensura,
Universidad Nacional de Rosario and Instituto de F\'{\i}sica
Rosario, Bv.\ 27 de Febrero 210 bis, 2000 Rosario, Argentina.}
\author{A.O.\ Dobry}
\address{Facultad de Ciencias Exactas Ingenieria y Agrimensura,
Universidad Nacional de Rosario and Instituto de F\'{\i}sica
Rosario, Bv.\ 27 de Febrero 210 bis, 2000 Rosario, Argentina.}
\author{D.C.\ Cabra}
\address{Laboratoire de Physique Th\'eorique, Universit\'e Louis
Pasteur, 3 rue de l'Universit\'e, F-67084 Strasbourg Cedex, France}
\address{Departamento de F\'{\i}sica, Universidad Nacional de la
Plata, C.C.\ 67, (1900) La Plata, Argentina} \affiliation{Facultad
de Ingenier\'\i a, Universidad Nacional de Lomas de Zamora, Cno. de
Cintura y Juan XXIII, (1832) Lomas de Zamora, Argentina.}
\author{T.\ Vekua}
\address{Laboratoire de Physique Th\'eorique, Universit\'e Louis
Pasteur, 3 rue de l'Universit\'e, F-67084 Strasbourg Cedex, France}
\address{Andronikashvili Institute of Physics, Tamarashvili 6,
0177 Tbilisi, Georgia}

\date{\today}

\begin{abstract}
We study the magnetic excitations on top of the plateaux states
recently discovered in spin-Peierls systems in a magnetic field. We
show by means of extensive density matrix renormalization group (DMRG)
computations and an analytic
approach that one single spin-flip on top of $M\!=\!1\!-\!\frac2N$ ($N\!=\!3,4,...$)
plateau decays into $N$ elementary excitations each carrying a
fraction $\frac1N$ of the spin. This fractionalization goes beyond the
well-known decay of one magnon into two spinons taking place on top
of the $M\!=\!0$ plateau. Concentrating on the $\frac13$ plateau ($N\!=\!3$) we
unravel the microscopic structure of the domain walls which carry
fractional spin-$\frac13$, both from theory and numerics. These
excitations are shown to be noninteracting and should be observable
in x-ray and nuclear magnetic resonance experiments.
\end{abstract}


\maketitle

Constitutive elements of condensed systems and their interaction
laws are all well known and in spite of this, modern condensed
matter physics is a field where new fundamental concepts arise
continuously, mainly due to strong correlation effects. The clue
is that the emergent laws governing a system of many
interacting bodies could have no direct relationship with the
behavior of each individual member. In other words, the interaction
processes could wash out the individual properties of the
constituents and give rise to excitations of fundamentally new
character.\cite{Anderson} Specifically, a new paradigm is now
arising in the field of strongly correlated electron systems, where
the concept of Fermi liquid theory is not applicable any more.
Collective excitations with quantum numbers essentially different
than those of the individual electrons are now predicted and
observed in a variety of systems.

The earliest example arose in the 1970s in the study of
conducting polymers as polyacetylene. For this system, it was
proposed that conduction was due to solitons carrying the
electronic unit charge but no spin. The emergence of these
quasiparticles carrying different quantum numbers than the
original constituents, was understood as a consequence of
electron-phonon interactions.\cite{poliacet} Another example
is provided by a two-dimensional layer of electrons in a high
magnetic field. In the so-called fractional quantum Hall effect
regime at a certain filling fraction corresponding to a plateau in
the conductivity, the charge of the elementary excitations is a
fraction of the electronic charge. The statistical properties of
these quasiparticles are intermediate between fermionic and
bosonic and they are termed ``anyons." \cite{QHE}

Understanding the mechanisms through which collective processes
could produce excitations different in character than the original
constituents of a solid state system, is currently under intense
study. In particular, making specific predictions of the effects
of these excitations on the experimental observations is a very
important issue of modern condensed matter physics.
\begin{figure}
\includegraphics[width=0.35\textwidth]{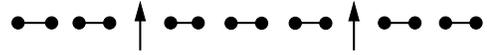}
\caption{\label{fig1ter}Deconfined spinons in dimerized chain, connected
black dots represent dimers - spin singlet combination of two neighboring
spins, arrow is for free spin.}
\end{figure}

In studying the properties of many-body systems, magnetic systems
have provided a fertile playground especially for elucidating very
important aspects of reduced dimensionality and strong correlations.
When one flips the spin of an individual electron (say
$S^z\!=\!-\frac12\rightarrow\!S^z\!=\!\frac12$) the total spin of
the system changes by one unity, $\Delta\!S^z\!=\!1$, so one has
created an excitation carrying spin $S^z\!=\!1$. In a
three-dimensional system this excitation was ascribed to be carried
by a bosonic particle called a magnon, a quantum of a spin wave. It
came as a big surprise when Fadeev and Takhtajan identified the
elementary spin quantum number of a spin wave as $S^z\!=\!\frac12$
in the one-dimensional world, calling them spinons.\cite{Fadeev} One
can ``have a look" at these spinons if one introduces sufficiently
strong frustration in the one-dimensional Heisenberg chain so that
the ground state becomes dimerized. Then the spinon acquires a
finite gap and it is visualized as a free spin separating the
different domains of dimerization as depicted in Fig. \ref{fig1ter}.

Apart from the natural $S^z\!=\!\frac12$ value (which could still be
ascribed to the individual electron) no other fractional values were
observed in experiments.

The present paper is devoted to the study of fractional spin
excitations that go beyond the usual fractionalization of a magnon
into two spinons discussed above. For example, as we discuss
below, the excitations on top of the $M\!=\!\frac13$ plateau carry a
fractional spin $S^z\!=\!\frac13$, which we dub ``tertions." This
fractionalization takes place due to collective effects in certain
magnetoelastic systems under a strong magnetic field. These
excitations should be observable in spin-Peierls systems like
CuGeO$_3$ and in Ising antiferromagnetic chains susceptible to
lattice deformations. These tertions should condense as a soliton
lattice in the ground state of a system under a magnetic field
greater than the value corresponding to the $M\!=\!\frac13$ plateau. This
would allow to directly observe these objects in nuclear magnetic resonance
(NMR), x-ray or neutron scattering experiments as has been the case with
closing the zero magnetization gap.\cite{kiri,fagot}

The lattice Hamiltonian of a frustrated spin chain coupled to
frozen phonons in a magnetic field reads as\cite{nosPRL}
\begin{eqnarray}
\label{eq:ham}
{\cal H}&=&\frac{1}{2}K \sum_{i} \delta_{i}^2
+ J_1\sum_{i} (1-A_1\delta_{i})\,
\vec{S}_{i} \cdot \vec{S}_{i+1} \nonumber \\
&+& J_2 \sum_i  \vec{S}_{i} \cdot
\vec{S}_{i+2}-H\sum_{i} S^z_i,
\end{eqnarray}
$H$ is measured in units where $g\mu_B\!=\!1$, $\delta_i$ is the
distortion of the bond between site $i$ and $i+1$, $K$ the spring
constant and the first term corresponds to the elastic energy loss.
$J_1$ sets the overall energy scale, and $\sqrt{\frac{J_1}{K}}$ a
corresponding distance scale. From now on, we fix $J_1\!=\!K\!=\!1$ to get
dimensionless energies and distances.

Recently we have shown that plateaux can be present for
magnetization values $M\!=\!1-\frac2N$, with $N\!=\!2,3,4,...$ being the
length of the periodic cell of the ground state in units of the
lattice constant. The actual presence of these plateau depends on
the strength of frustration, except for the $M\!=\!0$ plateau ($N\!=\!2$)
which is always present. The simplest nontrivial ones, at $M\!=\!\frac13$
($N\!=\!3$) and $M\!=\!\frac12$ ($N\!=\!4$), have been observed clearly in
numerical simulations for moderate values of frustration $J_2$ and
spin-lattice coupling $A_1$.\cite{nosPRL}

We have found that these plateaux are due to the next-to-leading
transfer processes becoming commensurate, in first order of the
spin-phonon interaction, and they appear at special rational
magnetization values in accordance with Ref. \onlinecite{OYA}. For the
$M\!=\!\frac13$ plateau this corresponds to the process of transferring two
particles from, say, the left to the right Fermi point, and for
the $M\!=\!\frac14$ plateau, a process involving the transfer of three
particles from the left to the right Fermi point. Those plateaux
are generically less wide than the zero magnetization plateau
which is caused by the doubling of the amplitude of the basic
transfer process at $M\!=\!0$.

Close to the zero magnetization plateau, the modulation of the
lattice distortions breaks into domains corresponding to a soliton
lattice.\cite{Brazovski} Domain walls carry spin $S^z\!=\!\frac12$ and
are deconfined. In analogy to the above picture our purpose is to
study the excitations on top of the nontrivial magnetization
plateaux at $M\!=\!1-\frac2N$, $N\!=\!3,4,...$ to show that one spin-flip
decays into $N$ free fractional spin excitations, with spin
$S^z\!=\!\frac1N$. To this end, we analyze the formation of a soliton
lattice on top of the $\frac13$ plateau state, which is an up-up-down
($uud$) modulated structure in the frustrated antiferromagnetic spin
chain coupled to adiabatic phonons.
\begin{figure}
\includegraphics[width=0.4\textwidth]{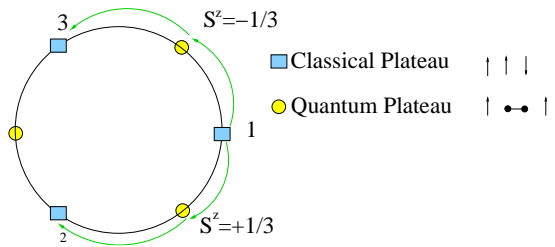}
\caption{\label{fig2ter}Values of the (periodic) bosonic field for
different $M\!=\!\frac13$ ground states. Squares correspond to the
three (classical) configurations and circles to the quantum
counterpart.\cite{Hida}}
\end{figure}

Fractionally charged excitations in the
systems with commensurability 3 were studied in the early 1980s in
one-dimensional electron-phonon systems numerically
\cite{SuSchrieffer} and by bosonization.\cite{Hara} In the case of 1/3
electron filling these works identified elementary excitations carrying
charge and spin values (in addition to polaronic excitation with ordinary
electronic quantum numbers):
$\Delta Q=\pm e/3$, $\Delta S=1/2$ and $\Delta Q=\pm 2/3$, $\Delta S=0$,
respectively. Our case corresponds to 1/2 electron filling, with completely frozen
charge fluctuations. As we will show in this case spin excitations will be
fractionalized in the units of $1/3$. Since the charge field is suppressed
there is no direct analogy between the quantum numbers of the excitations
for electronic systems and our magnetic system which
is equivalent to the system of spinless electrons.

We developed a self-consistent
harmonic approximation (SCHA) in analogy to the zero magnetization
case and our findings are fully confirmed by extensive DMRG
computations.

The bosonized version of Eq.(\ref{eq:ham}) reads like
\begin{equation}
\label{basic}
H=H_0+H_{ph}+H_{sp},
\end{equation}
where $H_0$ is a Gaussian part,
$H_{ph}$ is the adiabatic phonon part, and $H_{sp}$ is the spin-phonon
interaction term which, around $M\!=\!\frac13$ is given by
\cite{nosPRL}

\begin{equation}
-A_1\!\int\!dx\delta(x)[\beta:\cos(\sqrt{2\pi}\phi):+\gamma:\cos(2\sqrt{2\pi}\phi):],\label{hint1/3}
\end{equation}
where $\delta(x)$ is the smooth part of the displacement field in
the continuum limit and columns $:\cdots:$ indicate normal ordering of the vertex
operators with respect to
the ground state with magnetization $M$. For $k_F\!=\!\frac{\pi}{3}$ 
there are three inequivalent minima which
are degenerate, which correspond to the three different $uud$
arrangements. In terms of the phonon and bosonic fields, they
correspond to $\delta_p(x)\!=\!\delta_0 \cos[2k_F(x+p)]$ ($p\!=\!0,1,2$) and
$\sqrt{2\pi}\phi\!=\!0,\pm \frac{2\pi}{3}$, respectively. These three
structures are clearly observed in the numerical simulations.

An interesting observation is that singlets can always appear in
domain walls because when tunnelling from the first vacuum to the
second or third, the field rests on the intermediate pseudominimum in
between (which turns out to be a portion of the quantum
plateau indicated by yellow circles in Fig. \ref{fig2ter}).

Let us now analyze the excitations on top of the $M\!=\!\frac13$
plateau. In the pure spin case, the potential energy is given simply
by $V[\phi] \propto \int dx \cos\left({3 \sqrt{2\pi} \phi}\right)$,
which also has three degenerate minima \cite{LO}. From this
potential, one immediately concludes that the excitations on top of
the plateau correspond to massive kinks (whenever $V[\phi]$ is a
relevant perturbation) interpolating between these inequivalent
minima, and carry fractionalized spin-$\frac13$.\cite{LO}

In the spin-phonon case, the situation is more subtle, since now
the three minima correspond to combined magnetoelastic
configurations as we discussed above. To see how fractional spin
kinks arise in that case, we resort to a SCHA along the lines of
Refs. \onlinecite{DHN,NF,DI}

Following Refs. \onlinecite{DHN,NF,DI} we split $\phi$ into classical and
quantum components, $\phi\!=\!\phi_{c}+\phi_{q}$. Using the value of
$k_F$ for $M\!=\!\frac13$ and keeping only commensurate terms, we arrive
at the following potential for the classical bosonic field:
\begin{equation}
V[\phi_c]\sim -\int dx \cos\left({3\sqrt{2\pi}\phi_c}\right)
\end{equation}
which led us to conclude that kinks are similar to those in the
pure spin case, though now both spin and phonon modulations must
combine appropriately. Below we find the explicit expression for
the local magnetization and bond modulations and compare them with
our numerical results obtained by DMRG.

Let us start discussing how $\phi_c$ evolves as we walk around a
chain with periodic boundary conditions (PBC). We start from the vacuum
corresponding to $\sqrt{2\pi}\phi_c\!=\!0$, then we have a tunnelling
of $\sqrt{2\pi}\phi_c$ from $0\!\to\!\frac{2\pi}{3}$, at the position
of the first domain wall (let us call this point $x_1$), then a
tunnelling process from $\frac{2\pi}{3}\!\to\!\frac{4\pi}{3}$ takes
place (at $x_2$) and at the position of the third domain wall
($x_3$) the initial vacuum is restored by tunnelling
$\frac{4\pi}{3}\!\to\!0$. An analytic expression for $\phi_c$ can be
built up as a product of three soliton solutions of the
sine-Gordon model\cite{Rajaraman} centered at $x_1$, $x_2$ and
$x_3$. It reads
\begin{eqnarray}
\label{soliton}
3 \sqrt{2\pi} \phi_c(x)&=&\frac{1}{8\pi^2}\Big[4\arctan\{\exp[(x-x_1)/\xi]\}\nonumber\\
&&\times(2\pi+4\arctan\{\exp[(x-x_2)/\xi]\})\nonumber\\
&&\times(4\pi+4\arctan\{\exp[(x-x_3)/\xi]\})\Big]
\end{eqnarray}
with $\xi$ being the soliton width. From Eq.(\ref{soliton}) and
the bosonization formulas connecting $\phi(x\!=\!ia)$ with $S^z_i$
(Ref.\onlinecite{Giamarchi}) we extract the local magnetization of every three
sites,
\begin{eqnarray}
\label{Szanal}
 <\!S^z_{\alpha}(3x)\!>&=&\frac{1}{6 \pi}\partial_x\phi_c(x)-B_1\cos\left( \sqrt{2\pi}\phi_c(x)+\frac {4 \pi}{3}
 \alpha\right) \nonumber\\
 &-&B_2 \cos\left( 2\sqrt{2\pi}\phi_c(x)+\frac {8 \pi}{3} \alpha\right) +\frac 16 .
\end{eqnarray}

As anticipated, singlets indeed appear within domain walls. This is
because when tunnelling from one vacuum to another, the field passes
through the intermediate pseudominimum in between, which is exactly
a portion of the quantum plateau \cite{Daniel,Hida} (see Fig.
\ref{fig2ter}). Here we would like to note that in the absence of
the spin phonon coupling the 1/3 magnetization plaetau in the
$J_1-J_2$ model was for the first time identified for stronger
values of $J_2$ by Okunishi and Tonegawa\cite{OkunishiTonegawa1} who
also identified similar fractionalized spin-$1/3$ excitations around
it.\cite{OkunishiTonegawa2} They connected this excitation with a
domain wall in the Ising limit, microscopically different from a
singlet-core excitation that is realized in our case of additional
spin phonon interaction and SU(2) symmetric spin exchange.

We now undertake a numerical analysis of the lattice deformations
($\delta_i$) and the local magnetization ($\left<\!S^z_i\!\right>$)
around the plateau at $M\!=\!\frac13$. We have used an iterative
method based on a DMRG procedure to solve the adiabatic equation
corresponding to Hamiltonian (\ref{eq:ham}) along the lines stated
in Ref.\onlinecite{nosPRL}. To compare with the previous analytical
study, we consider PBC, and the calculations were carried out
keeping $m\!=\!200$ states, with a truncation error of order
$10^{-11}$ in the worst case.

\begin{figure}[hb]
\includegraphics[width=0.46\textwidth]{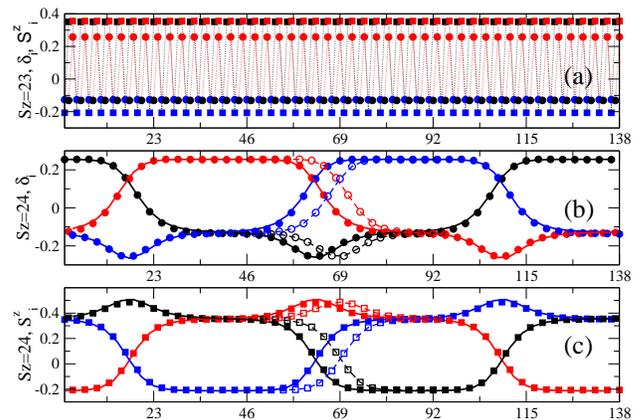}
\caption{\label{fig3ter}DMRG solution for $N_s\!=\!138,
J_2\!=\!0.5,$ and $A_1\!=\!0.6$. We represent with circles (squares)
symbols $\delta_i$ ($<\!S^z_i\!>$), using different colors for each
of the three sublattices. (a) Results
for $M=\frac13$ in our system. (b) and (c) show the results for
Solid lines in (b) and (c) correspond to the modulations obtained
within the SCHA (Ref.\onlinecite{NF}). Three well-defined
excitations are seen as what we call tertions. Open symbols and
dotted lines correspond to DMRG and bosonization results for a
second pattern where the central tertion is shifted. Both patterns
have the same energy, showing that tertions are noninteracting.}
\end{figure}

\begin{figure}[ht]
\includegraphics[width=0.45\textwidth]{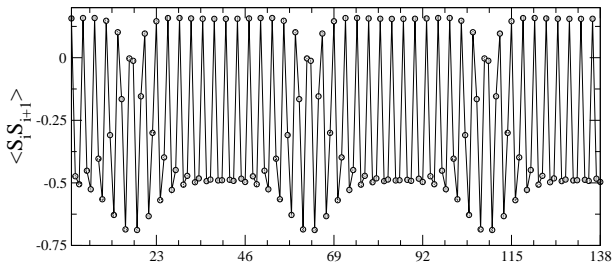}
\caption{\label{fig4ter}DMRG results for the same set of parameter
used in Fig. \ref{fig3ter} for the local spin-spin correlation
function.}
\end{figure}

With $M\!=\!\frac{2S^z_{\text{tot}}}{N_s}$, we want to study the
states for $M\!=\!\frac13$ and one unit of magnetization above it.
The iterative procedure for the $\delta_i$, takes around 100
iterations to achieve convergence in a particular $N_s$. Note
however that a periodic pattern with a wavelength
$\lambda\!=\!n\frac{2}{1+M}$ ($n$ integer) is expected for each
magnetization $M$.\cite{nosPRL} Therefore, to reduce the CPU time we
impose such a periodicity for both $M$, on the $\delta_i$ pattern in
our numerical calculation, choosing $N_s=138$. Then we study for
values of $S^z_{\text{tot}}=23$ and 24, and this enforcement help us
to obtain very accurate results for the states we are interested in.
In Fig. \ref{fig3ter} we show the results of $\delta_i$ and $S^z_i$
for a particular set of parameters where the plateau at
$M\!=\!\frac13$ is present. In Fig. \ref{fig3ter}(a), for
$S^z_{\text{tot}}\!=\!23, \lambda\!=\!3$ the up-up-down structure is
clearly seen, corresponding to a weak-weak-strong structure for the
bonds. Figures \ref{fig3ter}(b) and \ref{fig3ter}(c) show
magnetization and distortion respectively, for
$S^z_{\text{tot}}\!=\!24, \lambda\!=\!46$.

The patterns obtained for $\delta_i$ and $S^z_i$ are oscillatory on
the scale of the lattice constant. We separate the lattice in three
different sublattices to extract the smooth variations of the
relevant quantities. Three different excitations are clearly
identified which are characterized as domain walls of the $uud$
order. As the total spin of this state is $S^z\!=\!1$ above the
$M\!=\!\frac13$ state, each excitation carries $S^z\!=\!\frac13$ and
for this reason we term them tertions. Moreover, a very accurate
fitting could be found between the DMRG results and the analytic
form given in Eq. (\ref{Szanal}). Lines on Fig. \ref{fig3ter}(c)
were obtained from this expression with parameters $B_1\!=\!0.35$,
$B_2\!=\!0.03$, and the soliton width in units of three lattice
sites, $\xi\!=\!4.5$.

In Figs. \ref{fig3ter}(b) and \ref{fig3ter}(c), we added further
DMRG results and the analytical fitting, now shifting the position
of the second domain wall, and running the code again without
forcing the periodicity. The overall coincidence between the
bozonization and DMRG results and the fact that both states have the
same energy confirm that the excitations correspond to
noninteracting solitons with fractionalized spin $S^z\!=\!\frac13$.
DMRG results for other lattice sizes not shown here, lend further
support to this conclusion of independence, in particular since Eq.
(\ref{Szanal}) perfectly fits in all cases the numerical results
using the same set of constants $B_1, B_2$, and $\xi$. We also
checked that excitations behave similarly for different sets of
parameters where the $M\!=\!\frac13$ plateau is present.

Finally, let us  analyze the internal structure of these tertions.
Looking at the tertion placed at the center of the lattice, it can
be seen in Fig. \ref{fig3ter}(c) that $S^z$ has greater value at
position 63, and almost vanishes at sites 61-62 and 64-65. This fact
points towards singlet formation as we have predicted theoretically.
In fact, we have calculated the spin-spin correlation $\left<\!{\bf
S}_i\cdot {\bf S}_{i+1}\!\right>$ shown in Fig. \ref{fig4ter}, and
we obtain that the value $\sim\!-\frac34$ at the bonds around each
tertion is centered. Depending on the system size, the quantum
plateau portion can be longer or shorter.

\begin{figure}[hb]
\includegraphics[width=0.4\textwidth]{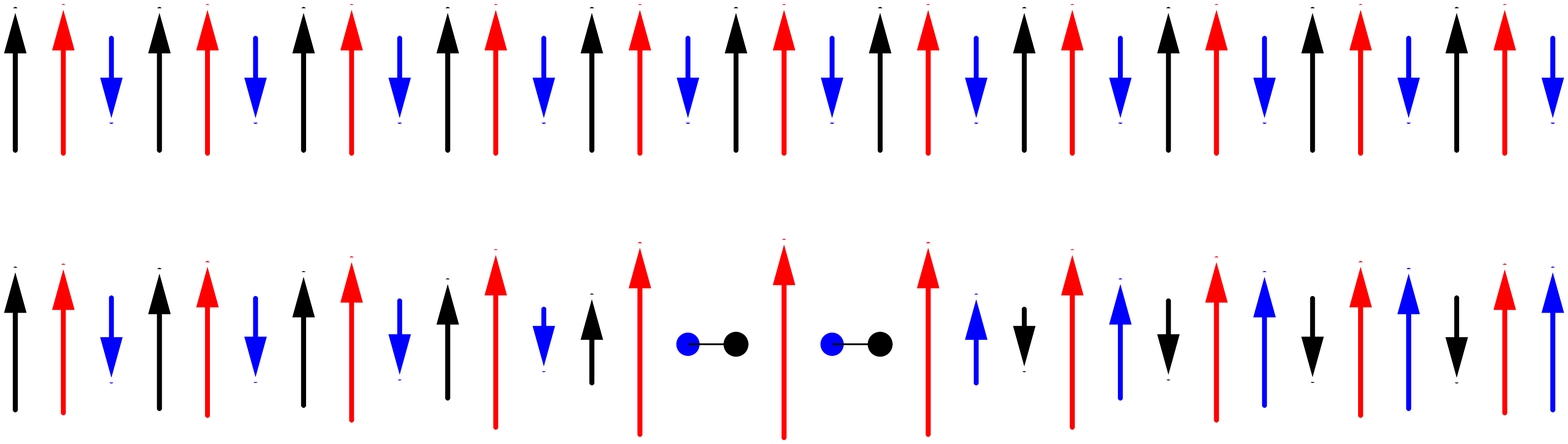}
\caption{\label{fig5ter}Scheme of the magnetic structure of the
state at $M\!=\!\frac13$ (upper panel) and a tertion (lower panel)
as given by DMRG calculation for the same parameters of Fig.
\ref{fig3ter}. The length of the arrows is proportional to
$<\!S^z_i\!>$.}
\end{figure}

In conclusion we have shown that plateaux in magnetoelastic systems,
independently of the mechanism that produce them, involve the
development of a soliton lattice at the threshold. We have also
shown that solitons-domain walls carry fractional spin values which
are generically smaller than $\frac12$, in particular for the
excitations around the $M\!=\!\frac13$ plateau, non-interacting
quasiparticles with fractional spin $S^z\!=\!\frac13$ arise. We have
also identified the core of the domain wall as singlets in the case
of the $M\!=\!\frac13$ plateau (Fig. \ref{fig5ter}). We hope that
our predictions will stimulate further high field experiments on
spin-Peierls compounds. Like for the case near $M\!=\!0$ in the
spin-Peierls material CuGeO$_3$,\cite{kiri} the lattice deformation
we predict could be measured in x-ray or neutron scattering
experiments. The local magnetic texture could otherwise be seen in
NMR experiments as in Ref. \onlinecite{fagot}.

The authors thank R.\ Santachiara, A.\ Tennant and G.\ Uhrig for
helpful discussions. This work was partially supported by ECOS-Sud
Argentina-France collaboration (Grant No. A04E03), PICS CNRS-CONICET
(Grant No. 18294), PICT ANCYPT (Grant No. 03-12409), and PIP CONICET
(Grant No. 5306).


\end{document}